\documentclass[12pt]{article}

\newcommand{\be}{\begin{equation}}
\newcommand{\ee}{\end{equation}}
\newcommand{\beqs}{\begin{eqnarray}}
\newcommand{\eeqs}{\end{eqnarray}}

\def\({\left(}
\def\){\right)}

\def\mxth{\mathsurround=0pt }
\def\xversim#1#2{\lower2.pt\vbox{\baselineskip0pt \lineskip-.5pt
x  \ialign{$\mxth#1\hfil##\hfil$\crcr#2\crcr\sim\crcr}}}

\renewcommand{\a}{\alpha}
\renewcommand{\b}{\beta}

\newcommand{\pa}{\partial}

\newcommand{\z}{\zeta}

\newcommand{\m}{\mu}
\newcommand{\n}{\nu}

\newcommand{\Ka}{{K\"ahler}}

\def\be{\begin{equation}}
\def\ee{\end{equation}}
\def\bea{\begin{eqnarray}}
\def\eea{\end{eqnarray}}

\newcommand{\ft}[2]{{\textstyle\frac{#1}{#2}}}

\newcommand{\eqn}[1]{(\ref{#1})}

\begin{document}
\begin{titlepage}
\begin{center}
\hfill ITP-UU-01/12  \\
\hfill SPIN-01/08  \\
\hfill YITP-01/18  \\
\hfill {\tt hep-th/0104215}\\
\vskip 20mm

{\Large {\bf Gauging Isometries on Hyperk\"ahler Cones \\[3mm]
and Quaternion-K\"ahler Manifolds }}

\vskip 10mm

{\bf Bernard de Wit$^{a}$, Martin Ro\v{c}ek$^b$  and
Stefan Vandoren$^b$ }

\vskip 4mm

$^a${\em Institute for Theoretical Physics} and {\em Spinoza
Institute}\\
{\em Utrecht University, Utrecht, The Netherlands}\\
{\tt  B.deWit@phys.uu.nl} \\[2mm]
$^b${\em C.N. Yang Institute for Theoretical Physics}\\
{\em SUNY, Stony Brook, NY 11794-3840, USA}\\
{\tt rocek@insti.physics.sunysb.edu}\\
{\tt vandoren@insti.physics.sunysb.edu}

\vskip 6mm

\end{center}

\vskip .2in

\begin{center} {\bf ABSTRACT } \end{center}
\begin{quotation}\noindent
We extend our previous results on the relation between 
quaternion-\Ka\ manifolds and hyperk\"ahler
cones and we describe how isometries, moment maps and scalar 
potentials descend from the cone to the quaternion-K\"ahler space.
As an example of the general construction, we discuss the gauging
and the corresponding scalar potential of hypermultiplets with the
unitary Wolf spaces as target spaces. This class includes   
the universal hypermultiplet.

\end{quotation}

\vfill
\flushleft{\today}

\end{titlepage}

\eject

\section{Introduction}

The scalar fields of $N=2$ hypermultiplets coupled to supergravity
parametrize a quaternion-K\"ahler (QK) manifold 
\cite{BagWit}. These manifolds appear as target spaces 
in the low-energy effective actions of type-II superstrings compactified on 
a Calabi-Yau three-fold, or heterotic strings compactified on K3.
The geometry of QK manifolds is best understood in terms of their 
hyperk\"ahler cones (HKC's) \cite{Swann,Galicki,DWKV,DWRV}.
This one-to-one correspondence is based on 
the $N=2$ superconformal quotient \cite{DWKV,DWRV}, and it can be 
exploited in a variety of applications, ranging from supersymmetry breaking 
in gauged supergravity, domain walls and supergravity flows, 
to quantum corrections to hypermultiplet moduli spaces of Calabi-Yau 
string compactifications.

In this paper, we discuss the gauging of isometries on QK spaces.
The strategy is to first gauge the corresponding isometries on the HKC,
where the computations are relatively easy, and subsequently 
project to the QK space. We construct the moment maps and the
resulting scalar potential, and explain how the gauge fields
of the isometries contribute to R-symmetry connections. 
As the scalars of the vector multiplets live on special K\"ahler
manifolds and contribute to the scalar potential, we review some
aspects of special K\"ahler geometry.

Finally, as a concrete example, we consider the unitary Wolf spaces, and in
particular, the universal hypermultiplet. We point out certain subtleties
in performing the QK quotient \cite{Galicki2} which can elegantly be avoided
by first taking a hyperk\"ahler quotient followed by an $N=2$ superconformal
quotient of the HKC.

We first summarize some of the results of \cite{DWKV,DWRV} about 
$4n$-dimensional HKC's and their corresponding $4(n-1)$-dimensional 
QK metrics. The HKC metric is determined
by a hyperk\"ahler potential $\chi$, satisfying
\be
D_A\partial_B \chi = g_{AB}\ , \qquad A=1,\ldots, 4n \ . \label{homothety}
\ee
The associated $N=2$ sigma model is therefore scale invariant, with a
homothetic conformal Killing vector $\chi^A=g^{AB}\chi_B$, where $\chi_A\equiv
\partial_A\chi$. Moreover, using the complex structures $\vec J$ of
the HKC, one can construct the Killing vectors associated with an
Sp(1) isometry group,
\be
\vec k{}^A = \vec J{}^{A}{}_{\!B}\, \chi^{B}\ .
\label{su(2)-vector}
\ee
This implies that the corresponding hypermultiplet action is rigidly $N=2$
superconformally invariant. 

The $N=2$ superconformal quotient amounts to gauging the superconformal
algebra (hence the sigma model couples to superconformal gravity),
and eliminating the dilatation and ${\rm Sp}(1)={\rm SU}(2)$ gauge
fields which are not part of the Poincar\'e supergravity multiplet.
The resulting sigma model has a quaternion-K\"ahler metric \cite{DWKV},
\be
G_{AB}=\frac{1}{\chi}\left(g_{AB}-\frac{1}{2\chi}\Big[ \chi_A\chi_B +
{\vec k}_A\cdot{\vec k}_B\Big] \right)\ ,
\label{quaternionhorizontal}
\ee
and quaternionic two-forms
\be
{\vec {\cal Q}}_{AB}=G_{AC}{\vec J}^C{}_B\ .\label{QK-forms}
\ee
A QK manifold also has an Sp(1) connection ${\vec {\cal V}}_A=
\chi^{-1}{\vec k}_A$. Its curvature is proportional to
the quaternionic two-forms\footnote{Since the curvature of the Sp(1)
connection is nonzero, we cannot trivialize the Sp(1) bundle. All 
quantities that tranform under Sp(1), such as ${\vec {\cal Q}}$, are 
defined on this bundle, and not just on the QK base space itself, and 
consequently are subject to local Sp(1) gauge transformations.},
\be
{\vec {\cal R}}_{AB}\equiv \partial _A{\vec {\cal V}}_B-\partial_B {\vec
{\cal V}}_A-{\vec {\cal V}}_A\times {\vec {\cal V}}_B=
-2{\vec {\cal Q}}_{AB}\ .
\ee
The constant on the right hand side is fixed by the normalization
of the metric \eqn{quaternionhorizontal}.

The metric and two-forms \eqn{quaternionhorizontal} and \eqn{QK-forms}
still carry indices in the $4n$-dimensional HKC, but they are
horizontal in the sense that they are orthogonal to the conformal Killing
vector $\chi^A$ and the Sp(1) Killing vectors ${\vec k}^A$. These vectors are
associated with a cone over an Sp(1) fibration of the QK space. The
fibration over the QK space is a $(4n-1)$-dimensional 3-Sasakian manifold.
To descend to the $4(n-1)$-dimensional QK space, we choose appropriate
gauge conditions corresponding to the scale and Sp(1) symmetries \cite{DWRV}.
This is most easily done by first choosing $2n$ holomorphic coordinates 
$z^a$ on the HKC for which $J^{3a}{}_b=i\delta^a{}_b$. 
The vector field $\chi^a$ is then holomorphic and one can single out 
a coordinate $z$ (with remaining coordinates $u^i,\ i=1,\ldots,2n-1$)
by defining
\be
\chi^a(u,z)\,{\partial\over\pa z^a} \equiv {\partial\over\partial z} \ .
\label{def-z}
\ee
In this basis, \eqn{homothety} can be solved:
\be
\label{chi-K}
\chi(u,\bar u; z,\bar z)={\rm e}^{z+{\bar z}+ K(u,{\bar u})}\ ,
\ee
where $K(u,\bar u)$ is the K\"ahler potential of a complex $(2n-1)$-dimensional
Einstein-K\"ahler space, which is the twistor space ${\cal Z}$ above the 
QK manifold \cite{Salamon,Swann}. The metric
\eqn{quaternionhorizontal} takes the following form in these coordinates,
\be
G_{i\bar\jmath} = K_{i\bar\jmath} - {\rm e}^{2K} X_i X_{\bar\jmath}\ .
\ee
All other components vanish consistent with horizontality.

In the coordinates $(u^i,z)$, the homothety and Sp(1) Killing vectors 
have components
\bea
&&\chi^a = -i k^{3a} = (0,\ldots,0,1) \ , 
\qquad \chi_a = i k^3_a= \pa_a
\chi =\chi (K_i,1) \ , \nonumber\\[2mm]
&&k^+_a = {\rm e}^{2z} (X_i(u),0)\ ,
\ \ \ \,\qquad\qquad k^+_{\bar a}=0 \ ,\qquad k^-_{\bar a}=(k^+_a)^* \ ,
\label{SU2-KV}
\eea
where $X_i(u)$ is a holomorphic one-form on the twistor space.
Its curl defines a holomorphic two-form $\omega_{ij}=-\partial_{[i}
X_{j]}$; the pair $(X,\omega)$ is called a contact structure.
The two-form $\omega$ lives in odd dimensions and has a unique (up to 
normalization) holomorphic zero-eigenvalue vector field 
$Y^i(u):\, \omega_{ij}Y^j=0$. Using $Y^i(u)$, we can distinguish 
a second special coordinate $\zeta$ (with remaining coordinates 
$v^\a,\ \a=1,\ldots,2n-2$) by defining
\be
Y^i(v, \zeta) \,{\pa\over\pa u^i}\equiv {\pa\over \pa \zeta} \ .
\label{defzeta}
\ee
To descend to the QK subspace, we adopt the gauge-fixing conditions $z=\z=0$. 
These conditions are imposed on all quantities of interest, such as 
the QK metric,
\be
\label{QK-metric}
G_{\alpha {\bar \beta}}=K_{\alpha {\bar \beta}}-{\rm e}^{-2K}X_\alpha
X_{\bar \beta}\ ,
\ee
the quaternionic structure,
\be
{\cal Q}^3_{\alpha \bar \beta}=-iG_{\alpha \bar \beta}\ ,\qquad
{\cal Q}^+_{\alpha \beta}={\rm e}^{-K}\Big(\omega_{\alpha \beta}+
2K_{[\alpha}X_{\beta]}\Big)\ ,
\ee
and the Sp(1) connection (${\cal Q}^-$ and ${\cal V}^-$ follow by complex
conjugation)  
\be
{\cal V}^3_\alpha=-iK_{\alpha}\ ,\qquad {\cal V}^+_\alpha={\rm e}^{-K}
X_{\alpha}\ .\label{Sp1-conn}
\ee

Our conventions are such that for compact QK manifolds, both
the HKC and QK metrics are positive definite. For the non-compact QK
spaces that one encounters in supergravity, the HKC metric $g$ is
pseudo-Riemannian (with a mostly minus signature), and $-G$ is
positive definite. This feature will be important in sect.~3.

\section{Isometries and moment maps}
\setcounter{equation}{0}
In this section, we show how triholomorphic isometries and their
moment maps descend from the HKC to the QK space. 
The isometries that commute with supersymmetry generate precisely the
group of triholomorphic isometries; they can 
be gauged by coupling the model to $N=2$ vector
multiplets \cite{ST,HKLR}. 
A triholomorphic Killing vector 
$k^A$ on a hyperk\"ahler manifold obeys ${\cal L}_k {\vec \Omega}=0$, where 
${\vec \Omega}_{AB}=g_{AC}{\vec J}^C{}_B$ denotes the triplet of
closed two-forms on the hyperk\"ahler space.
This implies there exists a triplet of moment maps $\vec \mu$, such that
\be
\vec \Omega_{AB} k^B=\partial_A \vec \mu \qquad
\Leftrightarrow \qquad -ig_{a\bar b} k^{\bar b}=\partial_a\mu^3
\ ,\quad \Omega_{ab}^+k^b=\partial_a \mu^+\ .\label{HK-MM}
\ee
By taking the covariant derivative of these equations, and using the 
quaternion algebra of complex structures, one shows that
the moment maps are harmonic functions on the hyperk\"ahler manifold:
\be
\Delta \vec \mu=0\ .\label{harm-MM}
\ee

HKC's also have Sp(1) isometries \eqn{su(2)-vector}; they are not 
triholomorphic but rotate the complex structures and supercharges. 
These isometries can only be gauged by coupling to the superconformal 
gravity multiplet, which contains Sp(1) gauge fields. 
A consistent gauging of both the triholomorphic and the Sp(1) isometries
requires them to mutually commute, and to commute with dilatations.
This leads to the identities \cite{DWKV} 
\be
\chi_A{\vec k}^A=0\ ,\qquad \chi_Ak^A=0\ ,\qquad
{\vec k}_Ak^A=-2{\vec \mu}\ . \label{kk-mu}
\ee
Note that the last equation defines the moment maps $\vec\mu$ algebraically in
terms of the Killing vectors; this is a special feature of HKC spaces
that has no counterpart on general hyperk\"ahler manifolds. 
Not surprisingly, it turns out that precisely these isometries descend 
to isometries of the underlying twistor and QK spaces.

A general analysis of the HKC Killing equation
for isometries that commute with the homothety and $k^3$ shows that
they are independent of the coordinate $z$, and take the form,
\be
k^i=-i\mu^i \ , \qquad k^z=i(K_i\mu^i-\mu)\ , \label{shk-isom}
\ee
where $\mu(u,\bar u)$ is a real function on ${\cal Z}$, with $\mu_i=\pa_i
\mu$ and $\mu^i=\mu_{\bar \jmath}K^{\bar \jmath i}$, satisfying
\be
D_i \pa_j \mu=0 \ . \label{Ddg=0}
\ee
This implies that the vector \eqn{shk-isom} is 
holomorphic.  The hyperk\"ahler potential
$\chi$ is invariant and $K(u,\bar u)$ changes by a \Ka\
transformation under the isometry, so that the twistor space $\cal Z$
admits an isometry generated by
\be
\label{Z-isom}
{k}_i=i\,\pa_i \mu\,,
\ee
and its complex conjugate. Note that $\mu$ is the K\"ahler moment 
map of the isometry on $\cal Z$; the case of constant $\mu$
corresponds to the $k^3$ isometry, which acts  
trivially on $\cal Z$. Because the HKC isometry \eqn{shk-isom} is
triholomorphic, there is an extra constraint on $\mu$, 
\be
\label{Lk-X}
{\cal L}_{k}X_i\equiv -i \mu^j\pa_j X_i-i \pa_i \mu^j\,X_j= -2i
(K_j\mu^j-\mu)X_i\ .
\ee

Using (\ref{shk-isom})-(\ref{Lk-X}) one shows that the moment map equations
\eqn{HK-MM} are solved explicitly by
\be
\mu^3(z,\bar z,u,\bar u)= \chi\, \mu (u,\bar u)\ ,
\qquad \mu^+(z,u)=\ft12 i\,{\rm e}^{2z}X_i(u)\, \mu^i(u)\ .
\ee
The integration constant is set to zero, since in a conformally invariant 
theory, the hyperk\"ahler moment maps must scale with weight two. 

The existence of moment maps on a generic hyperk\"ahler space is 
only guaranteed if the corresponding isometries are triholomorphic.
On a QK space, {\it all} isometries have moment maps. 
To see this, we introduce real indices ${\cal A},{\cal B}=1,\ldots,4(n-1)$ 
on the QK space, and define, for an {\it arbitrary} Killing 
vector \cite{Galicki2}
\be
\vec{\hat\mu}\equiv-\frac1{4(n-1)}\vec{\cal Q}^{\cal AB}D_{\cal A}
\hat k_{\cal B}\ .
\label{MM-KV}
\ee
It then follows, by taking the Sp(1) covariant derivative, that
\be
{\vec {\cal Q}}_{\cal A \cal B}\,{\hat k}^{\cal B}={\cal D}_{\cal
A}\,\vec{\hat \mu}\ . \label{QK-MM}
\ee
Here we have used the Sp(1) covariant constancy of the quaternionic
structure ${\cal D}_{\cal A} {\vec {\cal Q}}_{\cal B\cal C}
\equiv D_{\cal A} {\vec {\cal Q}}_{\cal B\cal C}-{\vec {\cal V}}_{\cal A}
\times {\vec {\cal Q}}_{\cal B\cal C}=0$, the
identity $D_{\cal A}D_{\cal B}{\hat k}_{\cal C}=R_{\cal B\cal C
\cal A \cal D}{\hat k}^{\cal D}$, which holds for any Killing 
vector, and the well known fact that 
the Sp(1) holonomy is equal to the Sp(1) curvature: $
R_{\cal ABC}{}^{\cal D} \,\vec{\cal J}^{\cal C}{}_{\cal D} =
2(n-1)\,\vec{\cal R}_{\cal AB}$.

Equation \eqn{QK-MM} is the generalization of the moment map equation for
hyperk\"ahler spaces \eqn{HK-MM}; it implies that the
Killing vector ${\hat k}$ rotates the quaternionic structure according to
\be
{\cal L}_{\hat k}{\vec {\cal Q}}= \left( 2{\vec {\hat \mu}}+
{\hat k}^{\cal A}{\vec {\cal V}}_{\cal A}\right) \times {\vec {\cal Q}}\ .
\label{LQ-rQ}
\ee
Hence, {\it any} isometry on a QK space has a triplet of moment maps.

Similar manipulations as those leading to
\eqn{harm-MM} yield \cite{Galicki2}
\be
{\hat \Delta} {\vec {\hat \mu}}=-4(n-1){\vec {\hat \mu}}\ ,\label{harm-QK-MM}
\ee
where ${\hat \Delta}$ is now the Sp(1) covariantized Laplacian. 
Recently, \eqn{MM-KV} and \eqn{harm-QK-MM} were discussed in 
\cite{cdkp} and \cite{AF}, respectively.

The HKC Killing vector $k$ descends to the 
QK space provided we add a compensating Sp(1) transformation to preserve 
the gauge $\zeta=0$. The isometry of the QK space is then, in the basis of
complex coordinates, \cite{DWRV}
\be
\label{Q-isom}
{\hat k}^\a=k^\a-\frac{k^\z}{X^\z}X^\a=-i \Big[ \mu^\a - \mu^\z
\,{X^\a\over X^\z}\Big] \ ,\qquad \hat k^\zeta=0\ ,
\ee
where $X^i=(X^\alpha,X^\zeta)=K^{i\bar \jmath}X_{\bar \jmath}$.
In the same coordinates, the moment map equations are
\bea
{\cal Q}^3_{\a\bar \beta}\, {\hat k}^{\bar \beta}&=&\partial_\alpha
{\hat \mu}^3 
-2i{\cal V}_\alpha^+\,{\hat \mu}^-\ ,\nonumber\\
{\cal Q}^+_{\alpha \beta}\,{\hat k}^\beta&=&\partial_\alpha {\hat
\mu}^+ +i({\cal V}^+_\alpha\, {\hat \mu}^3-{\cal V}_\alpha^3\, {\hat
\mu}^+)\ ,\nonumber\\ 
0&=&\partial_{\bar \alpha}{\hat \mu}^+-i{\cal V}^3_{\bar
\alpha}\,{\hat \mu}^+\ , 
\eea
and their complex conjugates. Using some of the identities 
in (3.17) from \cite{DWRV}, \eqn{Lk-X}, as well as the explicit form of 
the Sp(1) connection \eqn{Sp1-conn} and the quaternionic 
structure, we find the QK moment maps in terms of the twistor space
moment map,
\be
{\hat \mu}^3=\mu \ , \qquad {\hat \mu}^+=\ft12{i}\,{\rm e}^{-K}X_i\,\mu^i\ .
\label{Z-maps}
\ee
The relation between the HKC and QK moment maps is then simply 
\be
\mu^3=\chi\, {\hat \mu}^3\ ,\qquad \mu^+={\rm e}^{z-\bar z}\chi\,
{\hat \mu}^+\ .   \label{HK-QK-MM}
\ee
One can verify that a set of
HKC Killing vectors associated with a Lie algebra ${\bf g}$ induce
corresponding QK Killing vectors that generate the same algebra ${\bf g}$.

\section{Gauging and the scalar potential}
\setcounter{equation}{0}
We now turn to the scalar potential that arises after gauging the
isometries of a hypermultiplet action. This potential depends on the
spacetime dimension because the field content of the vector multiplets
associated with the gauging varies with the dimension. We focus
on the case of $d=4$, but generalizations 
to $d=5$ are straightforward. The idea is to start from the scalar
potential of the rigidly supersymmetric hypermultiplet action with a
HKC target space and to derive the corresponding potential after
coupling to supergravity; in this coupling the hypermultiplet target
space is converted into a QK space. 

On the HKC, the scalar potential arises after gauging triholomorphic
isometries by introducing $N=2$ vector multiplets coupled
superconformally to the hypermultiplets. We give the scalar kinetic
terms and the auxiliary field terms of the rigidly superconformally
invariant vector multiplet Lagrangian \cite{DWVP,DWLVP} (we use the
notation of \cite{DDKV}),   
\be 
{\cal L}= i({\cal D}_\m F_I\,{\cal D}^\m\bar X^I - {\cal D}_\m \bar
F_I\,{\cal D}^\m X^I ) +
\ft1{16} N_{IJ} \,\vec Y^I\cdot \vec Y^J\,.
\ee
We remind the reader that the action is expressed in terms of a
holomorphic function $F(X)$ in the complex scalar fields
$X^I$; superconformal invariance requires $F(X)$ to be homogeneous of
second degree.
The index $I$ labels the vector multiplets and subscripts
$I,J$ denote derivatives with respect to $X^I$,
$X^J$. Because the vector multiplet action itself plays an
ancillary role in what follows, we convert from special
coordinates to holomorphic sections only at the end of this section. 
The derivatives ${\cal D}_\m$ are gauge covariant derivatives
associated with a nonabelian gauge group. The scalars parametrize a
\Ka\ manifold with metric   
\be
N_{IJ}=-iF_{IJ}+i{\bar F}_{IJ}\ , \label{def-N}
\ee
(so that the scalar kinetic terms take the form $-N_{IJ} \,{\cal D}_\m
X^I\,{\cal D}^\m\bar X^J$).  
The vector multiplets contain SU(2) triplets of  auxiliary fields denoted
by $\vec Y^I$. For a nonabelian gauge group, the action also contains a
scalar potential that is quadratic in both $X^I$ and $\bar X^I$, which
we include below.

The hypermultiplets have a kinetic term and a potential equal to (see
{\it e.g.}  \cite{DWKV}, where one can also find the corresponding
expressions in terms of sections of the ${\rm Sp}(n) \times {\rm
Sp}(1)$ bundle; this is convenient when including the fermions),
\be
{\cal L} = -\ft12 g_{AB} {\cal D}_\m \phi^A \,{\cal D}^\m\phi^B
- 2g^2\,g_{AB}\, k^A_Ik^B_J\,X^I{\bar X}^J +\ft12 g \,\vec Y^I \cdot
{\vec \mu}_I\,, 
\ee
where $g$ denotes the coupling constant and $N^{IJ}$ is the inverse of
\eqn{def-N}. This form of the Lagrangian holds for arbitrary hyperk\"ahler
target spaces. In six
dimensions the $X^I\bar X^J$ term is absent because there are no scalar fields
in the vector multiplet. In five dimensions, the $X^I\bar X^J$
fields are replaced by the real scalar fields of the $d=5$ vector multiplet.
The potentials in various dimensions are of course related by dimensional
reduction \cite{ST}.

After eliminating the auxiliary fields of the vector 
multiplets we obtain the scalar potential, 
\be
{\cal L}^{\rm scalar}= -g^2\Big(2 g_{AB}\,k^A_Ik^B_J\,X^I{\bar X}^J+
N^{IJ}{\vec \mu}_I\cdot  
{\vec \mu}_J - N_{IJ} \,f^I_{KL} X^K\bar X^L \,f^J_{MN} X^M\bar X^N \Big)\ ,
\label{rigid-potential} 
\ee
where, for completeness, we include the potential for the nonabelian
vector multiplets; the structure constants of the nonabelian gauge
group are denoted by $f^I_{JK}$. Here and henceforth we assume that
the holomorphic function 
$F(X)$ is gauge invariant. The nonabelian gauge transformations define
holomorphic isometries of the \Ka\ space; their corresponding \Ka\
moment maps are equal to $\nu_I= f^J_{IK} (F_J\bar X^K + \bar
F_JX^K)$. Observe that when $g_{AB}$ and $N_{IJ}$ are positive
definite, the potential \eqn{rigid-potential} is nonnegative, as it
should be. 

When we couple the vector multiplets and hypermultiplets to conformal
supergravity, some of their component fields will act as compensating
fields and can be gauged away. Furthermore the ${\rm
SU}(2)\times {\rm U}(1)$ gauge fields of conformal supergravity are
integrated out. For the vector multiplets this implies that one is
taking a \Ka\ quotient upon which one obtains a {\it special} \Ka\
space; for the hypermultiplets the procedure amounts to performing the
$N=2$ superconformal quotient, so that the resulting target space is
QK. However, to end up with a Lagrangian that has the 
proper signs for the Einstein-Hilbert Lagrangian and for the
kinetic terms of the various fields, in particular for the scalars of
the vector multiplets and hypermultiplets, the \Ka\ and hyperk\"ahler
metrics that we start from cannot be positive definite. The
compensating scalar fields (two from the vector multiplets and four
from the hypermultiplets) should appear with opposite sign. 
With our conventions, the easiest thing is to change the overall
sign of the full Lagrangian. The resulting special \Ka\ and the QK
metrics are then negative definite. 

First we consider the kinetic terms of the combined Lagrangian. We
covariantize the derivatives with respect to the (bosonic) superconformal 
symmetries,
\bea
{\cal D}_\m X^I &=& \pa_\m X^I - (b_\m -iA_\m)X^I -
g\,f^I_{JK}\,W_\m^JX^K\,,\nonumber \\ 
{\cal D}_\m \phi^A &=& \pa_\m \phi^A - \chi^A \,b_\m + \ft14 \vec k^A
\cdot \vec V_\m - g \,k^A_I \,W_\m^I \,,
\eea
where $A_\m$ and $\vec V_\m$ are the gauge fields associated with ${\rm
SU}(2)\times {\rm U}(1)$ and $b_\m$ is the gauge field associated with
local scale transformations. Eliminating the gauge fields $A_\m$ and
$\vec V_\m$ by their field equations (explicit solutions will be
presented below) and setting $b_\m=0$ by a gauge transformation, one
obtains the result,
\bea
e^{-1} {\cal L}^{\rm kin}&=& N_{KL} X^K \bar X^L\; {\cal M}_{I\bar J}
\,{\cal D}_\m X^I\,{\cal D}^\m \bar X^J 
+\ft12 \chi\, G_{AB} {\cal D}_\m \phi^A\,{\cal D}^\m  \phi^{B} \nonumber\\
&&
-X^IN_{IJ} {\bar X}^J\,\Big[\ft16 R -\ft14 (\pa_\m\ln [X^IN_{IJ}
{\bar X}^J])^2 \Big] 
-\chi\,\Big[\ft16 R -\ft14 (\pa_\m \ln\chi)^2 \Big]\nonumber\\
&&+ D\Big[X^IN_{IJ} {\bar X}^J -\ft12\chi\Big] \,, \label{L-kin}
\eea
where the terms proportional to the Ricci scalar originate from the
covariantization with respect to conformal boosts and $D$ is an
auxiliary field which acts as a Lagrange multiplier and imposes the 
condition,
\be
\chi=2 X^IN_{IJ}\bar X^J\ . \label{D-constr}
\ee
The covariant derivatives depend only on the gauge fields $W^I_\m$ of
the vector multiplets. The horizontal metric associated with the
special \Ka\ space is
\be
{\cal M}_{I\bar J} = {1\over [N_{MN} X^M \bar X^N]^2} \Big[
N_{IJ}N_{KL} - N_{IK}N_{JL} \Big] \bar X^K X^L 
\ee

Subsequently we absorb the factor $\chi$ into the vierbein by
rescaling 
$e_\m^{\,a}\to \sqrt{2/\chi}\,e_\m^{\,a}$. Imposing the gauge
conditions for the QK target space, the result for \eqn{L-kin} reads, 
\be
e^{-1} {\cal L}^{\rm kin}= -\ft12 R+  {\cal M}_{I\bar J}
\,{\cal D}_\m X^I\,{\cal D}^\m \bar X^J 
+2 G_{\a \bar \b} {\cal D}_\m \phi^\a\,{\cal D}^\m \phi^{\bar \b}\,,
\label{L-kin2} 
\ee
where the derivatives are covariant with respect to the gauged
isometry group. Note that the fields $\phi^\a$ coincide with the QK
coordinates $v^\a$ introduced above \eqn{defzeta}.

The potential can now be evaluated straightforwardly, using the
results derived in sect.~2. Absorbing the factor $\chi$ into
the vierbein as before leads to the following expression for ${\cal
L}^{\rm scalar}$, 
\bea
e^{-1} {\cal L}^{\rm scalar}&=&  4  g^2\left[2\,G_{\alpha\bar \beta}\,
{\hat k}^\alpha_{(I}  {\hat k}^{\bar \beta}_{J)}  + 3\,{\vec {\hat
\mu}}_I\cdot {\vec {\hat \mu}}_J \right] \, 
{X^I{\bar X}^J \over N_{MN} X^M \bar X^N} 
\nonumber \\
&&+ g^2 \,N_{MN} X^M \bar X^N\; {\cal M}_{I\bar J}\, \left[ 4\, N^{IK}
N^{JL} {\vec {\hat \mu}}_K\cdot {\vec {\hat \mu}}_L -   {f^I_{KL}
X^K\bar X^L\over N_{PQ} X^P \bar
X^Q } \,{f^J_{MN} X^M\bar X^N\over N_{PQ} X^P \bar X^Q} \right] \,. 
\nonumber\\ 
&&~\label{V2}
\eea

The resulting potential is no longer positive definite. Using that
$G_{\a\bar\b}$ and ${\cal M}_{I\bar J}$ are negative definite, as
explained earlier, we see that the term proportional to $\vert
X^I\,\vec {\hat\m}_I\vert^2$ is negative whereas the others are
positive (note that $\chi$ must be positive). For an early discussion
of this, see {\it e.g.}, \cite{DWVP}. 

Our form of the potential \eqn{V2} is equivalent to earlier results
\cite{V-grav}, obtained by other methods. To appreciate the relation, we 
briefly discuss some aspects of 
special K\"ahler geometry (coordinatized by the vector multiplet 
scalars $X^I$). The Lagrangians \eqn{L-kin2} and \eqn{V2} do
not depend on the $X^I$ but rather on their ratios, and hence the
geometry is projective.
Consequently, we can replace the $X^I$ by 
holomorphic sections $X^I(z)$, defined modulo multiplication
by an arbitrary holomorphic function of the coordinates $z^{\hat I}$, {\it
i.e.},  $X^I(z) \to \exp[f(z)]\, X^I(z)$. Observe that this projective
invariance includes the U(1) invariance, for which we have, so far,
not imposed a gauge condition. At this point one may convert to the
formulation in terms of these holomorphic sections of the complex line 
bundle of special geometry \cite{strominger,CDF}. The number of independent  
holomorphic coordinates $z^{\hat I}$ is one less than the number 
of $X^I$, and this
difference can be traced back to the scale and U(1) invariance of the
original formulation. 
The special \Ka\ metric based on the coordinates 
$z$ has a \Ka\ potential\footnote{
  As is well known, the sections $X^I(z)$ can be combined with
  $F_I(z)$ into sections that transform covariantly under
  electric-magnetic duality. However, the duality is affected by the
  gauging, as can be seen from the fact that the various terms in the
  potential and in the moment maps (see below) are not symplectically
  invariant. Although we started from a holomorphic function $F(X)$, it
  is sufficient that only symplectic sections $(X^I(z),F_I(z))$
  exist in a way that is consistent with the requirements of special
  geometry \cite{CRTVP}. }
\be 
{\cal K}(z,\bar z) = - \ln \Big[ i \bar X^I(\bar z) \,F_I(X(z)) -i\bar
F_I(\bar X(\bar z)) \, X^I(z) \Big] \,,
\ee
which, under the projective transformations, changes by a \Ka\
transformation: ${\cal K}\to {\cal K} - f(z) -\bar f(\bar z)$. 
The potential \eqn{V2} is then expressed in terms of projective
invariants, such as $X^I(z)\,\bar X^J(\bar z)\,\exp [{\cal K}(z,\bar
z)]$, and in terms of the (inverse) special K\"ahler metric
${\cal K}^{{\hat I}{\hat {\!\bar J}}}$ via the relation 
(see, {\it e.g.} \cite{DW})
\be
N^{KL}={\rm e}^{\cal K}\left[{\cal K}^{{\hat I}{\hat {\!\bar J}}}
(\partial_{\hat I}+\partial_{\hat I}{\cal K})X^K(z)\,
(\partial_{\hat {\!\bar J}}+\partial_{\hat {\!\bar J}}{\cal K}){\bar X}^L
(\bar z) - X^K(z){\bar X}^L(\bar z)\right]\ .
\ee

We note that for the hypermultiplet sector there also exists a
formulation in terms of local sections, in this case of an ${\rm
Sp}(1)\times {\rm Sp}(n)$ bundle. The existence of this associated
quaternionic bundle is known from general arguments \cite{Swann} and
was explained for supersymmetric models in \cite{DWKV}.

Let us briefly return to the expressions for the ${\rm SU}(2)\times
{\rm U}(1)$ gauge fields, which pick up certain corrections associated
with the gauging. These corrections show up in the coupling to
the gravitini, which are related by supersymmetry to the fermion
mass terms. The corresponding expressions read, 
\bea 
A_\m &=& -\ft12 [N_{MN} X^M \bar X^N]^{-1} ( \bar F_I
\stackrel{\leftrightarrow} {\pa}_\m  X^I - \bar
X^I \stackrel{\leftrightarrow} {\pa}_\m  F_I) + g\, W^I_\m \,\hat\nu_I
\,,\nonumber \\ 
\vec V_\m &=& -2 \,\pa_\m\phi^{\cal A} \,\vec {\cal V}_{\cal A}
 -4 g\,W_\m^I\,\vec {\hat\m}_I  \,,
\eea
where we have imposed the gauge conditions on the HKC
and used the definition for the special \Ka\ moment map, $\hat \n_I =
\n_I [N_{MN} X^M \bar X^N]^{-1}$. Observe that this expression is
indeed consistent with the projective invariance associated with the
sections $X^I(z)$ (but it is {\it not} consistent with
electric-magnetic duality). The HKC and QK moment maps can be
expressed explicitly in terms of the associated bundles. The
relevant expressions are given in eqs. (3.35) and (5.7) of \cite{DWKV}.

\section{The unitary Wolf spaces}
\setcounter{equation}{0}
As an example we construct the moment maps for the QK unitary
Wolf spaces, 
\be
X(n-1)=\frac{{\rm U}(n-1,2)}{{\rm U}(n-1)\times {\rm U}(2)}\ .
\ee
These spaces have real dimension $4(n-1)$. The universal hypermultiplet 
corresponds to $n=2$; moment maps and gaugings of abelian isometries 
were discussed in four dimensions in \cite{d=4gauged-UHM}, whereas 
a more general gauging in five dimensions appeared in 
\cite{d=5gauged-UHM,cdkp}. Our construction not only gives a more uniform
treatment of all moment maps, but is also applicable for $n\geq 2$ and other
QK spaces.

As explained in \cite{DWRV}, the HKC above $X(n-1)$
can be  constructed from a hyperk\"ahler quotient \cite{Hitchin} 
of ${\bf C}^{2n+2}$
with respect to a U(1) isometry which acts with opposite phases on the
homogeneous coordinates $z_+^M,\,z_{-M}$, with $M=1,\ldots, n+1$, of 
${\bf C}^{2n+2}$. On these coordinates there is a {\it linear} action of 
${\rm SU}(n-1,2)$:
\be
z_+^M\rightarrow U^M{}_{\!N}\,z^N_+ \ ,\qquad 
z_{-M}\rightarrow (U^{-1})^N{}_{\!M}\,z_{-N} \ .
\ee
Clearly, these transformations define triholomorphic isometries.
The reality condition on the ${\rm SU}(n-1,2)$ matrices is 
\be
{\bar U}^{\bar M}{}_{\!\bar N}=\eta^{\bar M 
M}(U^{-1})^N{}_{\!M} \,\eta_{N\bar N}\ ,
\ee
where $\eta_{M\bar N}={\rm diag}(-\cdots -++)$.

The U(1) hyperk\"ahler quotient imposes the holomorphic moment map
condition $z_{-M}z_+^M=0$; we solve this in a specific gauge:
\be
z_+^{n+1}=1\ , \qquad z_{-n+1}=-z^a_+z_{-a}
\ .\label{cond}
\ee
One then finds the hyperk\"ahler potential for the HKC of $X(n-1)$ \cite{DWRV}:
\be
\chi_{2n}= 2\chi_+\chi_-\ ,
\ee
where
\be
\chi_+={\sqrt {\eta_{M\bar N}z_+^M
{\bar z}_+^{\bar N}}}\ , \qquad \chi_-={\sqrt {\eta^{M\bar N}
z_{-M}{\bar z}_{-\bar N}}}\ .
\ee

The triholomorphic SU($n-1$,2) transformations on the HKC above $X(n-1)$
were given in \cite{DWRV}; in infinitesimal form, with $U={\bf 1} + 
i\theta^I T_I;\, I=1,\ldots,n(n+2)$, they determine the HKC 
Killing vectors $(\delta z_+^M,\delta z_{-M})=\theta^I 
(k_{I\,+}^M,k_{I\,-M})$,
\bea
k_{I\,+}^M&=&i(T_I)^M{}_{\!N}\,z_+^N-iz_+^M
(T_I)^{n+1}{}_{\!N}\,z^N_+\ ,\nonumber\\
k_{I\,-M}&=&-i(T_I)^N{}_{\!M}\,z_{-N}+iz_{-M}
(T_I)^{n+1}{}_{\!N}\,z^N_+\ .
\eea
Notice the compensating terms needed to preserve the conditions
\eqn{cond}. 

The moment maps on the HKC now simply follow from \eqn{kk-mu} 
and \eqn{SU2-KV} (with the holomorphic one-form $k^+={\rm e}
^{2z}X=2z_{-M}{\rm d}z_+^M$):
\bea
\mu_I^3&=&\frac{2i}{\chi}\Big[\chi_-^2(\bar z_+^{\bar M}\eta_{N\bar M}k_{I+}^N)
+\chi_+^2(\bar z_{-\bar M}\eta^{\bar M N}k_{I-N})\Big]\ ,\nonumber\\
&=&-\frac{2}{\chi}\Big[\chi_-^2(z_+^M(T_I)^N
{}_{\!M}\,\eta_{N\bar M}\,{\bar z}_+^{\bar M})
-\chi_+^2(z_{-M}(T_I)^M{}_{\!N}\,\eta^{N\bar M}
{\bar z}_{-\bar M})\Big]\ ,\nonumber\\
\mu_I^+&=&-z_{-M}k^M_{I+}\ = \ -iz_{-M}(T_I)^M{}_{\!N}\,z_+^N\ ,\label{HKC-MM}
\eea
subject to the constraints \eqn{cond}. 

Finally, we determine the moment maps and Killing vectors on $X(n-1)$; we
set 
\be
z_{-n}={\rm e}^{2z} \ ,\qquad z_{-i}={\rm e}^{2z}w_i \ ,\qquad
2z_+^n=\zeta\ , \qquad z_+^i=v^i\ ,
\ee
where $a=(i,n)$ with $i=1,\cdots ,n-1$. The coordinates on 
$X(n-1)$ are then $v^\alpha=\{w_i,v^i\}$. The moment maps  
follow from the HKC moment maps \eqn{HKC-MM} using
\eqn{HK-QK-MM}. The QK Killing vectors are obtained from \eqn{Q-isom}, 
so we must first compute the compensating Sp(1) transformation 
associated with $X^\alpha/X^\zeta$. These follow from the K\"ahler
potential on the  twistor space
\be
K( v,w, \zeta,  \bar v,\bar w, \bar\zeta) = \ln \left[ 2
\,\chi_+(v,\zeta,\bar v,\bar \zeta)  \;\chi_-(v,w,\zeta,\bar v, \bar
w,\bar \zeta)\right] \ ,
\ee
with
\be
\chi_+= \sqrt{1 +\ft14  \zeta\bar \zeta +\eta_{i\bar \jmath}\,v^i
\bar v^j}\ ,\qquad
\chi_- = \sqrt{1+ \vert \ft12 \zeta+ v^iw_i\vert^2 +\eta^{i\bar \jmath }
w_i\bar w_j} \ ,
\ee
where $\eta_{i\bar \jmath}=-\delta_{i\bar \jmath}$. Applying the general 
formula (4.16) from \cite{DWRV} yields
\be
\frac{X^{w_i}}{X^\zeta}=-\ft12\eta_{i\bar \jmath}\,{\bar v}^j\,
\frac{\chi_-^2}{\chi_+^2}\ ,\qquad 
\frac{X^{v^i}}{X^\zeta}=\ft12 (\eta^{i\bar 
\jmath}{\bar w}_j+\bar w_k\bar v^k v^i)\ .
\ee
Moreover, using (henceforth we suppress the index $I$ for the generators)
\bea
k^\zeta|_{z=\zeta=0}&=&2i\left[T^n{}_{\!{n+1}}+T^n{}_{\!i}\,v^i\right]
\ ,\\[3mm]
k^z|_{z=\zeta=0}&=&\ft12i\left[T^{n+1}{}_{\!n}\,(w_iv^i)-T^n{}_{\!n}
-T^i{}_{\!n}\,w_i+T^{n+1}{}_{\!{n+1}}+T^{n+1}{}_{\!i}\,v^i\right]\ ,\nonumber
\eea
we get the following expression for the QK Killing vectors:
\bea
{\hat k}^{v^i}&=&\left[iT^i{}_{\!j}-iv^iT^{n+1}{}_{\!j}\right]v^j+
iT^i{}_{\!{n+1}}-iv^iT^{n+1}{}_{\!{n+1}}-\ft12 k^\zeta 
\Big(\eta^{i\bar \jmath}\bar w_j+v^i 
\bar w_k\bar v^k\Big)\ ,\nonumber\\[5mm]
{\hat k}^{w_i}&=&-\left[iT^j{}_{\!i}-iv^jT^{n+1}{}_{\!i}\right]
w_j-iT^n{}_{\!i}+w_i\left[iT^{n+1}{}_{\!{n+1}}+iT^{n+1}{}_{\!j}\,v^j\right]
\nonumber\\
&&-2k^z\,w_i+\frac{\chi_-^2}{2\chi_+^2} k^\zeta \, \eta_{i\bar \jmath}\,
\bar v^j \ .
\eea

As a specific illustration, we gauge a sample isometry for the universal
hypermultiplet. Its QK metric is given by
\bea
G_{w{\bar w}}&=&-\, \frac{1- v{\bar v}}{2\,[1- w\bar w(1 - v\bar
v)]^2}\ , \nonumber\\[2mm]
G_{w\bar v}&=&\frac{\bar w v}
{2\,[1- w\bar w(1 - v\bar v)]^2} \ ,  \nonumber\\[2mm]
G_{v\bar v}&=&-\,\frac{1- w\bar w(1- v\bar v)^2}
{2\,[1- w\bar w(1- v\bar v)]^2[1- v\bar v]^2}\ .\label{UH-metric}
\eea
We choose the generator
\be
T=\ft13 \pmatrix {1 & 0 & 0 \cr 0 & -2 & 0 \cr 0 & 0 & 1}\ .
\label{gen1}
\ee
On the HKC we then find, 
\be
k^z=\ft12 i\ ,\qquad k^\zeta=-i\zeta\ ,\qquad k^v=0\ ,\qquad 
k^w=-iw\ ,
\ee
with corresponding moment maps
\be
\mu^3=\frac{1}{\chi}\left[\ft12\chi_-^2\,\zeta\bar \zeta
-2\chi_+^2\,{\rm e}^{2(z+\bar z)}\right]\ ,\qquad
\mu^+=\ft12i\,{\rm e}^{2z}\zeta\ .
\ee

Because $k^\zeta|_{\zeta=0}=0$ in this case, the QK Killing vector is 
\be
{\hat k}^v=0 \ ,\qquad  {\hat k}^w=-iw \ .
\ee
The fact that this vector is an isometry of the metric \eqn{UH-metric}
can easily be checked by direct calculation. The corresponding QK
moment maps are  
\be
{\hat \mu}^3=-\ft12 \frac{1}{[1-w\bar w(1-v\bar v)]}\ ,\qquad
{\hat \mu}^+=0\ .
\ee
Note that, due to the gauge choice $\zeta=0$, ${\hat \mu}^+$ vanishes on 
the QK space, but not on the HKC. This is not surprising; because the
QK moment maps are sections of the Sp(1) bundle, one can always employ
a local Sp(1) transformation to make ${\hat \mu}^{\pm}$ vanish. This
is in contrast  
with triholomorphic isometries on hyperk\"ahler spaces, where  
the Sp(1) bundle is trivial and a vanishing moment map 
would imply that the Killing vector vanishes. This 
observation clarifies the supersymmetric low-energy effective dynamics
of the Higgs branch (at scales below the masses of the gauge fields
$W_\mu^I$): when some of the moment maps vanish, one might worry that the
QK quotient \cite{Galicki2} becomes degenerate.
However, since we can lift everything to the HKC where this problem
does not arise, it is guaranteed that this will not happen.
Therefore, the geometry of the light
scalar-field manifold is best understood from the perspective of the HKC.

\noindent 

\vspace{8mm}

\noindent
{\bf Acknowledgement}\\
We would like to thank Jan Louis and Toine Van Proeyen for stimulating
discussions. This work was done with partial support from NSF 
grant No.\ PHY-9722101.

%

\end{document}